\documentclass[namedreferences]{SolarPhysics}
\usepackage[optionalrh,solaenum]{spr-sola-addons} 
\usepackage{graphicx}                    
\usepackage{amssymb}                     
\usepackage{color}                       
\usepackage{url}                         

\begin{document}

\begin{article}

\begin{opening}

\title{Further Evidence Suggestive of a Solar Influence on Nuclear Decay Rates}

%
\author{P.A.~\surname{Sturrock}$^{1}$\sep
        E.~\surname{Fischbach}$^{2}$\sep
        J.H.~\surname{Jenkins}$^{3,2}$      
       }

%
\runningauthor{Sturrock, Fischbach \& Jenkins}

%
  \institute{$^{1}$ Center for Space Science and Astrophysics, Stanford University, Stanford, CA 94305-4060, USA. 
                     email: \url{sturrock@stanford.edu}\\ 
             $^{2}$ Department of Physics, Purdue University, West Lafayette, IN 47907, USA
                     email: \url{ephraim@purdue.edu} \\
             $^{3}$ School of Nuclear Engineering, Purdue University, West Lafayette, IN 47907, USA
                     email: \url{jere@purdue.edu} \\
             }

\begin{abstract}
Recent analyses of nuclear decay data show evidence of variations suggestive of a 
solar influence. Analyses of datasets acquired at the Brookhaven National Laboratory (BNL) 
and at the Physikalisch-Technische Bundesanstalt (PTB) both show evidence of an annual 
periodicity and of periodicities with sidereal frequencies in the neighborhood of 12.25 
year$^{-1}$ (at a significance level that we have estimated to be 10$^{-17}$). It is notable that 
this implied rotation rate is lower than that attributed to the solar radiative zone, 
suggestive of a slowly rotating solar core.  This leads us to hypothesize that there may 
be an ``inner tachocline'' separating the core from the radiative zone, analogous to 
the  ``outer tachocline'' that separates the radiative zone from the convection zone. The 
Rieger periodicity (which has a period of about 154 days, corresponding to a frequency 
of 2.37 year$^{-1}$) may be attributed to an r-mode oscillation with spherical-harmonic 
indices $l=3, m=1$, located in the outer tachocline. This suggests that we may test 
the hypothesis of a solar influence on nuclear decay rates by searching BNL and PTB data 
for evidence of a ``Rieger-like'' r-mode oscillation, with $l=3, m=1$, in the inner 
tachocline. The appropriate search band for such an oscillation is estimated to 
be 2.00-2.28 year$^{-1}$. We find, in both datasets, strong evidence of a periodicity 
at 2.11 year$^{-1}$. We estimate that the probability of obtaining these results by 
chance is 10$^{-12}$.

\end{abstract}

%

\end{opening}

%
\section{Introduction}

\inlinecite{jen09}, \inlinecite{fis09}, and \inlinecite{Jav10} have analyzed data from measurements of the decay rates of $^{32}$Si and $^{36}$Cl acquired at the Brookhaven National Laboratory (BNL; \inlinecite{alb86}) and from measurements of the decay rate of $^{226}$Ra acquired at the Physikalisch-Techniche Bundesanstalt (PTB; \inlinecite{Sie98}), finding evidence of small but statistically significant annual variations in those rates. These findings led those authors to suggest that nuclear decay rates may be influenced (via some unknown mechanism) by the Sun. These analyses have been criticized by \inlinecite{Coo09}, \inlinecite{Nor09}, and \inlinecite{Sem09}, but detailed responses to these criticisms have been presented by \inlinecite{Jen10} and by \inlinecite{Jav10}.

The possibility of a solar influence on decay-rate variations has led us to search for periodicities that might be indicative of solar rotation. Our recent analyses (\opencite{Stu10a}, \citeyear{Stu10b}) of BNL and PTB data yield evidence in both datasets of a periodicity with a frequency in the neighborhood of 11.25 year$^{-1}$.\footnote{An earlier analysis \cite{Stu10b} led to the estimate 11.93 year$^{-1}$ for BNL, but our recent analysis \cite{Stu10a} shows that this estimate arose from the convolution of two separate peaks in the power spectrum, one at 11.24 year$^{-1}$ and the other at 13.08 year$^{-1}$.}   The standard shuffle test \cite{Bah91} indicates that there is a probability of only 10$^{-17}$ of finding by chance both peaks at the same frequency in a search band of 10-15 year$^{-1}$. We have also examined the data by means of a new test, called the ``shake'' test \cite{Stu10b}, which yields results very similar to those obtained by the shuffle test.

This periodicity is suggestive of the influence of a core or the environment of a core that has a sidereal rotation frequency of about 12.25 year$^{-1}$. [The synodic rotation frequency---the frequency as observed from Earth---is lower than the sidereal (absolute) rotation frequency by 1 year$^{-1}$.] The thin region at normalized radius approximately 0.7 that separates the radiative zone from the convection zone is known as the ``tachocline,'' since it is the location of a sharp change in rotation rate. If there is a similar sharp change in rotation rate between the core, with sidereal rotation frequency at or close to 12.25 year$^{-1}$, and the radiative zone, with a mean sidereal rotation rate of about 13.7 year$^{-1}$ \cite{Sch98}, there would be a second or ``inner'' tachocline, which could conceivably have properties similar to those of the ``outer'' tachocline.

There is supporting evidence for this conjecture. A combined analysis (\opencite{Stu08}, \citeyear{Stu09}) of low-energy solar neutrino data from the Homestake (\opencite{Cle98}, \opencite{Dav96}) and GALLEX (\opencite{Ans93}, \citeyear{Ans95}) experiments and of total solar irradiance data from the ACRIM experiment (\opencite{Wil79}, \citeyear{Wil01}) yields evidence of a periodicity of approximately 11.85 year$^{-1}$, corresponding to a sidereal rotation frequency of 12.85 year$^{-1}$.  Also, a recent analysis of Mount Wilson solar diameter measurements yields evidence of oscillations originating in a region with sidereal rotation frequency of 12.08 year$^{-1}$ \cite{Stu10}.  These inferred rotation frequencies are not the same, but they are both lower than current estimates of the rotation rate of the radiative zone \cite{Sch98}, so that they could both originate in a tachocline.

The well-known solar Rieger periodicity (which was originally discovered in gamma-ray flare data; \opencite{Rie84}) has a period of about 154 days, corresponding to a frequency of approximately 2.37 year$^{-1}$. We have proposed that this periodicity may be attributed to an r-mode oscillation that occurs in the known tachocline \cite{Stu99,Stu06,Stu09}, an interpretation that receives strong support from the recent article by \inlinecite{Zaq10}.

These considerations suggest that we may test the hypothesis of a solar influence on nuclear decay rates as follows. If there is indeed an inner tachocline, it may be the seat of a Rieger-like oscillation at a frequency determined by the band of rotation frequencies of that region.  This leads to a specific proposal that we present in Section 2. Our search for this periodicity is presented in Section 3. We discuss the results in Section 4.

\section{Proposal}

r-mode oscillations \cite{Pap78,Pro81,Sai82,Wol86} are retrograde waves in a rotating fluid with frequencies determined, to good approximation, in terms of $l$ and $m$ (two of the three spherical harmonic indices) and the absolute or ``sidereal'' rotation frequency, $\nu_{R}$ . The values of $l$ and $m$ are restricted by

\begin{eqnarray}
l=2,3,\dots{},~~m=1,2,\dots{},l
\label{eq:1}
\end{eqnarray}

The r-mode frequencies, as they would be measured by an observer co-rotating with the Sun, are given to good approximation by

\begin{eqnarray}
\nu{}\left(l,m\right)=\frac{2m\nu_{R}}{l\left(l+1\right(}
\label{eq:2}
\end{eqnarray}

Equation (2) yields the 2.37 year$^{-1}$ Rieger frequency for the combination $l=3,m=1$ and $\nu_{R}=14.22 \rm{yr}^{-1}$, while other values of $l,m$ and $\nu_{R}$ and are disfavored. We find, from estimates of the internal solar rotation rate obtained from helioseismology \cite{Sch98}, that the Sun has this rotation rate at normalized radius 0.71, placing it in the tachocline.

Taking into account (a) the inferred sidereal rotation rate of 12.25 year$^{-1}$ with an uncertainty of 0.13 year$^{-1}$ derived from BNL and PTB data, (b) the estimated sidereal rotation rate of 12.85 year$^{-1}$ derived from solar neutrino and total solar irradiance data, and (c) the estimated sidereal rate of 12.08 year$^{-1}$ derived from Mount Wilson diameter data, we adopt 12.0 year$^{-1}$ for the lower limit of the sidereal rotation frequency of the inner tachocline.  For the upper limit, we adopt the mean rotation rate of the radiative zone, 13.7 year$^{-1}$. Adopting the same values of l and m $(l = 3, m = 1)$ as in our interpretation of the Rieger oscillation, we see from Equation (2) that the corresponding search band for the frequency of a Rieger-type periodicity associated with the hypothetical inner tachocline is 2.00--2.28 year$^{-1}$.

These considerations lead to the following proposal: \textit{The hypothesis of a solar influence on nuclear decay rates leads us to expect possible periodicities in BNL and PTB data in the range 2.00--2.28 year$^{-1}$.}

\section{Analysis}

We show in Figures 1 and 2 sections of the power spectra recently derived from BNL and PTB datasets, respectively (\opencite{Stu10a}, \citeyear{Stu10b}). The complete BNL dataset comprises 366 measurements in the time interval 1982.11 to 1989.93. We point out that, due to a change in the apparatus in 1986, \inlinecite{alb86} utilized only data from 1982 to 1986 in their determination of the $^{32}$Si half-life. However, we have chosen to analyze the complete dataset in our power spectrum analysis, since the slight discontinuity in the dataset should not affect the frequency content (although it would affect the determination of the $^{32}$Si half-life, which was the objective of the BNL experiment).

 \begin{figure} 
 \centerline{\includegraphics[width=0.7\textwidth,clip=]{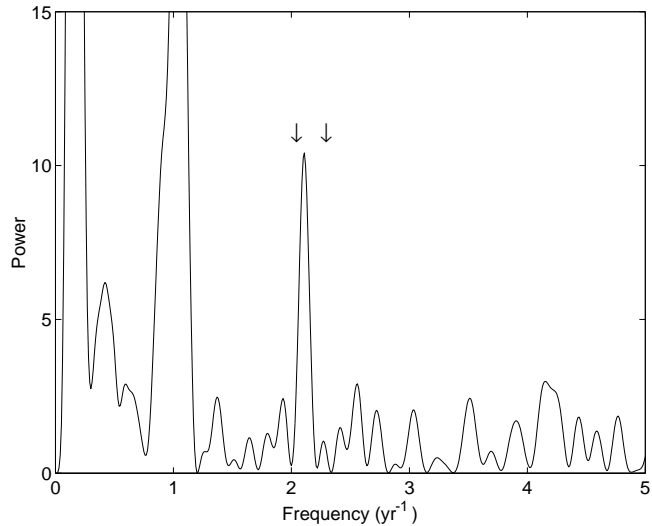}}
 \caption{Power spectrum formed from the BNL data. As mentioned in the text, this figure incorporates additional data obtained between 1986 and 1990, which were not utilized by Jenkins et al. (2009) or by Fischbach et al. (2009). The arrows indicate the search band 2.02 year$^{-1}$ to 2.28 year$^{-1}$. The peak is found at $2.107\pm0.004$ year$^{-1}$ with power $S$ = 10.09.}\label{fig:?}
 \end{figure}

 \begin{figure} 
 \centerline{\includegraphics[width=0.5\textwidth,clip=]{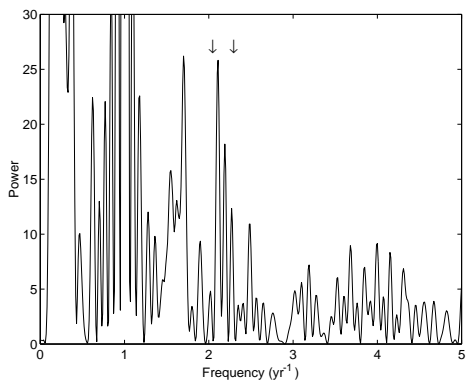}}
 \caption{Power spectrum formed from the PTB data. The arrows indicate the search band 2.02 year$^{-1}$ to 2.28 year$^{-1}$. The peak is found at  with power S = 25.83.}\label{fig:?}
 \end{figure}

Following the advice of the BNL experimenters, we have examined the ratio of the measurements of the decay rates of 32Si and 36Cl, each detrended to take account of the mean decay rate. We also examine the 1,966 PTB measurements of the decay rate of 226Ra, similarly detrended. Power spectra were formed using a likelihood procedure \cite{Stu05a} that is equivalent to the Lomb-Scargle procedure \cite{Lom76,Sca82}. The power spectra are complex, showing strong features at 1 year$^{-1}$ and at other frequencies (such as the neighborhood of 1.7 year$^{-1}$ in Figure 2). We plan to discuss these low-frequency features in a future article, but we here focus on features within the search band.

The BNL power spectrum shows a peak at 2.11 year$^{-1}$ with power $S1$ = 10.09. In order to obtain an error estimate, we adopt the convention of determining the ``1-sigma'' values. For a normal distribution of a quantity, the 1-sigma error is the displacement that leads to an increase in the probability of the null hypothesis by the factor $e^{0.5}$. In power-spectrum analysis, the probability of the null hypothesis (that the peak is due to normally distributed random noise) is given by $e^{-S}$ \cite{Sca82}.  Hence the 1-sigma offset is that at which $S$ is decreased by 0.5. In order to determine this offset for the BNL power spectrum, it is necessary to use a frequency resolution of only 0.001 year$^{-1}$. We then obtain the estimate $2.108\pm{}0.012$ year$^{-1}$ for the location of the peak. 

A similar analysis of the PTB data yields a peak at 2.11 year$^{-1}$ with power $S_2{}=25.83$.  Adopting a frequency resolution of 0.001 year$^{-1}$, we obtain the estimate $2.107\pm{}0.005$ year$^{-1}$ for the location of the peak. Taking into account the error estimates, the frequencies of the BNL and PTB peaks are clearly compatible.

We may examine the correlation between these two power spectra by forming the joint power statistic \cite{Stu05b} given, to good approximation, by

\begin{eqnarray}
J=\frac{1.943X^2}{\left(0.650+X\right)}
\label{eq:5}
\end{eqnarray}

where $X$ is the geometric mean of $S_1$ and $S_2$:

\begin{eqnarray}
X=\left(S_1{}S_2\right)^{1/2}
\label{eq:6}
\end{eqnarray}

This statistic has the property that if $S_1$ and $S_2$ are each distributed exponentially (as is the case for a time series derived from normally distributed random noise; \opencite{Sca82}), then $J$ also is distributed exponentially. We show $J$ as a function of frequency in Figure 3. The peak at 2.11 year$^{-1}$ has the value $J$ = 30.65. On taking account of our recent discussion of the false-alarm probability of peaks in power spectra \cite{Stu10}, we estimate that probability to be $1.3\times{}10^{-13}$. On adopting the finer frequency resolution, we find that the frequency of the peak value of the JPS is $2.107\pm{}0.005$ year$^{-1}$. 

 \begin{figure} 
 \centerline{\includegraphics[width=0.7\textwidth,clip=]{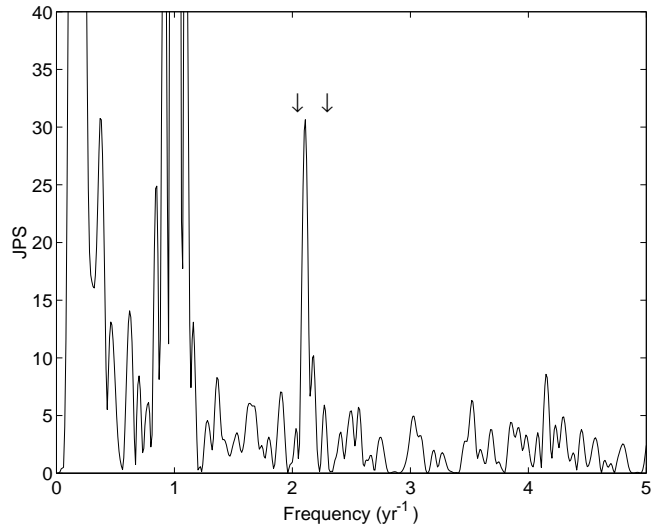}}
 \caption{The joint power statistic formed from the BNL and PTB power spectra. The arrows indicate the search band 2.02 year$^{-1}$ to 2.28 year$^{-1}$. The peak is found at $2.107\pm0.005$ year$^{-1}$ with joint power statistic $J$ = 30.65.}\label{fig:3}
 \end{figure}

 \begin{figure} 
 \centerline{\includegraphics[width=0.7\textwidth,clip=]{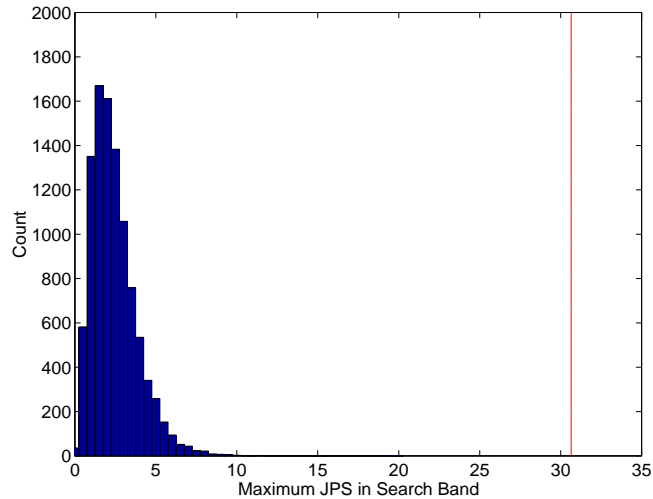}}
 \caption{Histogram formed from 10,000 shuffle simulations of the joint power statistic formed from the BNL and PTB power spectra. The vertical red line indicates the value (30.65) of the joint power statistic for the peak at 2.11 year$^{-1}$.}\label{fig:4}
 \end{figure}

Although the separation between peaks is given approximately by $1/T$, where $T$ is the duration of the dataset, we have seen that the precision of an individual peak depends critically on the peak power, and can be much less than the $1/T$ value. The frequency estimates derived from BNL data, from PTB data, and from the joint power statistic all differ significantly from 2.00 year$^{-1}$, the frequency of the first harmonic of the annual oscillation.

The significance of a peak in the power spectrum and the properties of the joint power statistic are based on an assumed exponential form for the distribution function of the power, which is appropriate if the non-signal measurements (the ``noise'') conform to a normal distribution \cite{Sca82}.  We examine the validity of this assumption for the BNL and PTB datasets in the Appendix, where we show that mapping the measurements onto a strictly normal distribution has a negligible effect on the power spectrum, from which we may infer that any departure of the measurements from a normal distribution has no significant adverse effect on our analysis.

However, in order to obtain a robust significance estimate, we have applied the shuffle test \cite{Bah91} that does not depend on any assumed form of the power distribution function. We have computed $J$ for 10,000 Monte Carlo simulations generated by the shuffle procedure, shuffling both datasets and noting the maximum value of $J$ in the frequency band 2.00--2.28 year$^{-1}$ for each simulation. The results from the shuffle test are shown in histogram form in Figure 4 and in a logarithmic display in Figure 5. A projection of the results shown in Figure 5 indicates that the probability of obtaining by chance a value of $J$ as large as the actual value (30.65) is about 10$^{-12}$, slightly more conservative than the false-alarm probability.

 \begin{figure} 
 \centerline{\includegraphics[width=0.7\textwidth,clip=]{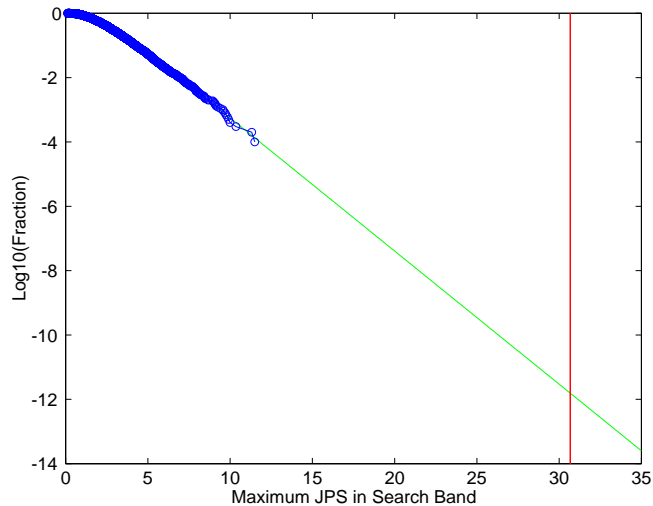}}
 \caption{Logarithmic display of the results of 10,000 shuffle simulations of the joint power statistic formed from the BNL and PTB power spectra. The projection of this curve indicates that the probability of obtaining by chance a value of $J$ as large as the actual value (30.65) is about 10$^{-12.}$}\label{fig:5}
 \end{figure}

We have also carried out simulations using the shake test \cite{Stu10b}, which has the advantage over the shuffle test that it is less sensitive to low-frequency modulations and secular trends. However, the results of the shake and shuffle tests prove to be virtually identical, indicating that there is negligible probability of obtaining by chance a value of the JPS as large as or larger than the actual value. We find, therefore, that this analysis is consistent with our proposal.

As a further precaution against the possibility that the oscillations at 2.11 year$^{-1}$ represent the first harmonic of the strong annual oscillations, we have carried out the following test. We have formed a ``template'' of the actual annual variation of the BNL data, and divided the actual measurements by the corresponding values of the template. This procedure should remove all effects of the annual oscillation, including harmonics. We find that the peak at 2.11 year$^{-1}$ remains in the power spectrum. We have applied the same procedure to the PTB dataset with the same result. This test indicates that the peaks at 2.11 year$^{-1}$ in the BNL and PTB power spectra may not be attributed to a harmonic of the annual oscillation.

\section{Discussion}
\textit{•}
It would appear that this study further strengthens the case for a solar influence on nuclear decay rates. It does not offer further insight into the physical mechanism of this influence, but it supports the inference that the origin of the influence is to be found below the radiative zone---either in or near the core.

The results of our analysis are therefore relevant to the question of the rotation rate of the solar core. It has to date been difficult to determine this quantity by means of helioseismology, and the results have been conflicting:  \inlinecite{Cha99} reported evidence for a downturn in the rotation rate below normalized radius 0.15, but \inlinecite{Cha01} later estimated the rate to be close to that of the radiative zone. On the other hand, \inlinecite{Gar07} have estimated that the rate may be three to five times faster than that of the radiative zone. \inlinecite{How09}, in her recent review of solar rotation, concluded that ``the rotation of the inner core …remain(s) unclear.'' The results presented in this article provide further evidence (in addition to \inlinecite{Stu10a,Stu10b}) that rotation of the solar core is slower than that of the radiative zone.  However, we are in a very early stage of learning about the structure and rotation of the deep solar interior. Any model that we can propose at this time will no doubt need to be revised in the future.

One may ask why r-mode oscillations should be preferentially excited in a tachocline. \inlinecite{Zaq10}, who also attribute the Rieger periodicity to an r-mode oscillation in the outer tachocline, have shown that r-modes become unstable in a layer where there is both a toroidal magnetic field and \textit{latitudinal} differential rotation, and that this instability favors $m = 1$ modes. However, this interpretation does not explain why the instability should be confined to the tachocline, since similar conditions may be presumed to exist throughout the convection zone. Our current conjecture is that r-modes are unstable in the presence of \textit{radial} differential rotation. We plan to investigate this hypothesis in the near future. We plan also to search for evidence of other r-modes, in addition to the mode with $l = 3, m = 1$, which offers an explanation of the Rieger oscillation.


%

%

%
 \appendix
 
\section{The Rank-Order Normalization (RONO) Procedure}

The power-spectrum analyses in this article have been carried out using a likelihood procedure \cite{Stu05a} that is equivalent to the Lomb-Scargle procedure \cite{Lom76,Sca82}. However, these procedures are based on the assumption that the background (non-signal) data have a normal distribution.  Since there is no reason to believe that the background data for the BNL and PTB experiments are distributed strictly in this way, we here examine the question of whether the departures of the BNL and PTB datasets from normal distributions are of any consequence. We address this question by the following procedure, which we refer to as the ``RONO,'' for ``Rank-Order NOrmalization,'' procedure.

The dataset comprises a sequence of measurements, $x_n, n=1,\dots{},N$, which we denote by $\left\lbrace x \right\rbrace$, taken at times $t_n, n=1,\dots,N$, which we denote by $\left\lbrace t \right\rbrace$. We now arrange the sequence of measurements in ascending order. We denote the re-arranged measurement sequence by $\left\lbrace \tilde{x} \right\rbrace = R \left\lbrace x \right\rbrace$.

\begin{eqnarray}
f=\mathop{erf} \left( g \right) , g=\mathop{erfinv} \left( f \right)
\label{eq:7}
\end{eqnarray}

We now define $\left\lbrace \tilde{y} \right\rbrace$ by

\begin{eqnarray}
\tilde{y}=\mathop{erfinv} \left( f_n \right)
\label{eq:8}
\end{eqnarray}

where

\begin{eqnarray}
f_n= \frac{n}{N+1}, n=1,\dots,N.
\label{eq:9}
\end{eqnarray}

 \begin{figure} 
 \centerline{\includegraphics[width=0.7\textwidth,clip=]{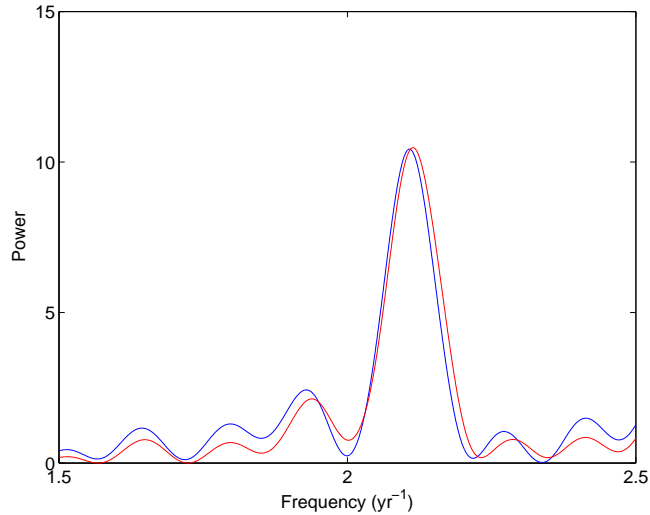}}
 \caption{The power spectrum formed from BNL data, after the application of the RONO procedure, is shown in red. The original power spectrum is shown, for comparison purposes, in blue. The original peak had power 10.09 at frequency 2.11 year$^{-1}$. The power spectrum derived from the normalized data has power 10.43 at the same frequency 2.11 year$^{-1}$.}\label{fig:6}
 \end{figure}

We have applied this transformation to the BNL data, and then applied the likelihood power spectrum analysis to the normalized data $\left\lbrace y \right\rbrace$. The result is shown in Figure 6, in which the original power spectrum (derived from the measurements $\left\lbrace x \right\rbrace$) is shown in blue, and the new power spectrum (derived from the normalized measurements $\left\lbrace y \right\rbrace$) is shown in red. The normalized data lead to a peak at 2.11 year$^{-1}$ with peak power 10.43. We see that the frequency of the peak is indistinguishable from the value in the original analysis, and the peak power is changed only slightly (from 10.09). 

The same operation on the PTB data leads to a power spectrum that has a peak at 2.11 year$^{-1}$ with power 25.50---again no change in frequency, and only a slight change in power from the original estimate of 25.83.

%
\begin{acks}
We are indebted to D. Alburger and G. Harbottle for supplying us with the BNL raw data, and to H. Schrader for supplying us with the PTB raw data. The work of PAS was supported in part by the NSF through Grant AST-06072572, and that of EF was supported in part by U.S. DOE contract No. DE-AC02-76ER071428.
\end{acks}

%

%

\end{article} 
\end{document}